\documentclass[letter]{aa}

\usepackage{txfonts}
\usepackage{graphicx}
\usepackage{color}
\usepackage{natbib}
\usepackage{rotating}
\usepackage{url}
\bibpunct{(}{)}{;}{a}{}{,} 

\begin{document}

\title{Probing the origin of the microwave anomalous foreground}
\author{N. Ysard
 \and M.A. Miville-Desch\^enes
  \and L. Verstraete}
\offprints{Nathalie Ysard, \email{nathalie.ysard@ias.u-psud.fr}}
\institute{Institut d'Astrophysique Spatiale, UMR8617, Universit\'e Paris-Sud, F-91405, Orsay, France}
\abstract
{The Galactic anomalous microwave emission detected between 10 and 90 GHz is a major foreground to CMB fluctuations. Well correlated with dust emission at 100 
$\mu$m, the anomalous foreground is interstellar but its origin is still debated. Possible carriers for this emission are spinning, small dust grains
carrying a permanent electric dipole.}
{To probe the origin of the anomalous foreground, we compare microwave data to dust IR emission on an angular scale of 1\degr, and search for specific signatures 
predicted by models of spinning dust.}
{For the anomalous foreground, we use the 23 GHz all-sky map deduced from WMAP data by Miville-Desch\^enes and collaborators. The infrared dust emission is traced by IRAS 
data. Models show that spinning dust emission is little sensitive to the intensity of the radiation field ($G_0$) for $10 \lesssim \nu \lesssim 30$ GHz, while the mid-IR emission 
produced by the same small dust grains is proportional to $G_0$. To test this behaviour in our comparison, we derive $G_0$ from the dust temperature maps of Schlegel and 
collaborators.}
{From all-sky maps, we show that the anomalous foreground is more strongly correlated with the emission of small grains (at 12 $\mu$m) than with that of large grains (at 100 $\mu$m). 
In addition, we show that the former correlation is significantly improved when the 12 $\mu$m flux is divided by $G_0$, as predicted by current models of spinning dust. 
The results apply to angular scales greater than 1$^\circ$. Finally, from a model fit of the anomalous foreground, we deduce physical properties for PAHs that are in good 
agreement with those deduced from mid-IR spectroscopy.}
{}
\authorrunning{}
\titlerunning{}
\maketitle

\section{Introduction}

As part of an effort towards accurate measurements of CMB fluctuations, experiments have motivated a detailed study of the Galactic foregrounds in the GHz-range. 
\citet{Kogut}, \citet{Leitch}, and \citet{Oliveira97} found an unexpected emission excess between 10 and 90 GHz, which is correlated with dust far-IR but not 
with synchrotron emission. To avoid an inaccurate interpretation, this excess has been referred to as an {\em anomalous foreground}. Low frequency observations have shown 
that it has a rising spectrum for $\nu \lesssim 30$ GHz \citep{Oliveira1999,Banday2003,Finkbeiner2004,Davies2006}. Both this behaviour and also the flux level of the excess 
are incompatible with what is known about the usual Galactic components in this spectral range: synchrotron, free-free, and thermal dust emission \citep{Oliveira1999, Oliveira04, 
Lagache, Finkbeiner2004, Miville2008}. If the anomalous foreground is caused by spinning, small grains as proposed by Draine \& Lazarian (1998, hereafter DL98) it should 
correlate more strongly with the mid-IR emission of small grains \citep{Oliveira2002} than with the large grain (BG) far-IR emission. Until now, there is only incomplete evidence that 
this is the case, the difficulty being the subtraction of zodiacal light in mid-IR data. From a comparison of WMAP data to HI data, \citet{Lagache} showed that the anomalous foreground 
rises with decreasing column density, in a similar way to the emission of small, transiently heated grains. Towards the dark cloud L1622, \citet{Casassus2006} found a stronger spatial 
correlation between the CBI 31 GHz and the IRAS 12 and 25 $\mu$m bands than with the 60 and 100 $\mu$m bands. In addition, models predict that the spinning dust emission is little 
sensitive to the intensity of the radiation field for $10 \lesssim \nu \lesssim 30$ GHz (DL98, Ali-Ha{\"i}moud et al. 2009, Ysard \& Verstraete 2009 hereafter YV09). \citet{Casassus2008} 
tested this result in the $\rho$ Oph molecular cloud: by comparing mid-IR (IRS, 13.3 $\mu$m) and 31 GHz (CBI) data, they showed that they are well correlated if the 
former is corrected for the exciting radiation field. In this paper, we further test the above predictions of spinning dust models by comparing, on a 1 degree scale, enhanced IRAS data 
(IRIS) and the anomalous all-sky map of Miville-Desch\^enes et al. (2008, hereafter MD08). We also derive some physical properties of dust and compare them to those deduced 
from IR spectroscopy.

\noindent The paper is organized as follows. Section 2 describes the observational predictions of spinning dust models and how they can be used to probe the origin of the 
anomalous foreground. Section 3 presents the data sets used to reach this goal. Section 4 shows how the anomalous foreground correlates with dust 
emission. Section 5 lists some remarkable fields and presents the type of information that we expect to derive from the anomalous foreground study. Finally, Section 6 presents 
our conclusions.

\section{Behaviour of the spinning dust emission}

\label{section_spinning_dust_emission}

\begin{figure}[!ht]
\resizebox{\hsize}{!}{\includegraphics{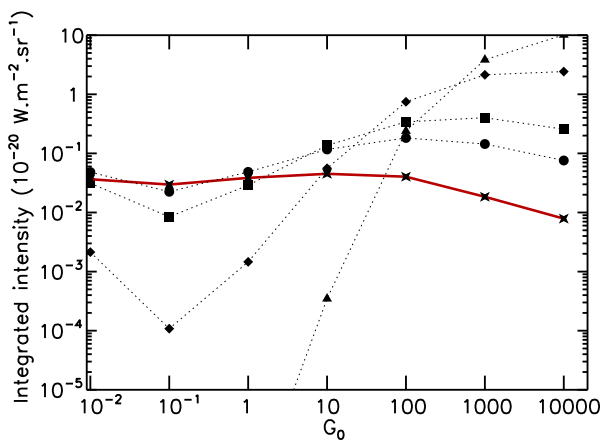}}
\caption{Results obtained with the model of YV09, for a MRN size distribution ($N_C = 30-216$), where $N_C$ is the number of C-atoms in the PAH and we assume a proton 
density of $n_H=$ 30 cm$^{-3}$. Other gas phase parameters (temperature, and electron, proton, and C$^+$ abundances) were computed with CLOUDY 05.07 \citep{CLOUDY} in the 
optically thin limit and assuming 130 (320) ppm of carbon (oxygen) in the gas phase.  Integrated intensity in the WMAP bands vs $G_0$: stars (red line) show the 23 GHz band, 
circles the 33 GHz band, squares the 41 GHz band, diamonds the 61 GHz band, and triangles the 94 GHz band.}
\label{model_spinning} 
\end{figure}

Nanometric-sized grains or Polycyclic Aromatic Hydrocarbons (PAHs) emit mostly in the mid-IR, whereas large grains dominate the FIR emission. The PAH emission is known to scale with the 
intensity of the radiation field $G_0$\footnote{Scaling factor for the radiation field integrated between 6 and 13.6 eV. The standard radiation field corresponds to $G_0=1$ 
and to an intensity of $1.6\times 10^{-3}$ erg/s/cm$^2$ \citep{Parravano2003}.} \citep{Sellgren1985}. This is also true for the emission of large grains and for $G_0<100$. We show in 
Fig. \ref{model_spinning} the behaviour of the spinning dust emission with $G_0$ in photometric bands predicted by our model (YV09). Models predict that the spinning dust emission 
near 23 GHz is almost independent of $G_0$ when $0.01\leq G_0\leq 100$ (Ali-Ha{\"i}moud et al. 2009, YV09, Fig. \ref{model_spinning}). This has strong observational consequences. 
If anomalous foreground {\it is} spinning dust emission, we expect there to be a stronger correlation between anomalous foreground at 23 GHz and IR emission divided by $G_0$ than with 
IR emission alone. Moreover this correlation should be stronger for 12 $\mu$m than 100 $\mu$m IRAS bands because the former traces the emission of small grains.

\section{Data sets}

To carry out these correlations, we need maps of the anomalous and dust emission as well as for the $G_0$ values. IRAS is a natural data set to study dust IR emission. Our 
IR template is the new generation of IRAS images, called IRIS \citep{IRIS}, in the 12 and 100 $\mu$m bands, which is corrected for most of the remaining instrumental problems 
of the IRAS/ISSA dataset. Point sources were removed in the IRIS plates (at 5 arcmin resolution) using the method described in \citet{IRIS}. The plates were then 
projected onto the Healpix grid, where an ecliptic-oriented filtering was applied to remove residual zodiacal light emission (Miville-Deschenes et al. in preparation). Finally, the IRIS 
all-sky maps were convolved with a 1 degree FWHM Gaussian, smoothing out any imperfections related to the point source subtraction.

\noindent \citet{Miville2008} performed a separation of components in the WMAP bands, using a physical approach to describe the Galactic foregrounds. We use this anomalous template at 
23 GHz, inferred from their ``Model 4''. The main assumption made to obtain this map is that polarized emission at 23 GHz is dominated by synchrotron (no assumption about any 
correlation with dust). 

\noindent Finally, the $G_0$-map is deduced from the BG temperature map of \citet{Schlegel1998} that is inferred from the 140/240 DIRBE ratio. We assume that the interstellar 
radiation field has the same spectral distribution as the standard field of \citet{ISRF}, everywhere in the Galaxy, and that the BG spectral index is $\beta = 2$ \citep{Draine1984}. 
The energy balance of a single grain of size 0.1 $\mu$m then yields $G_0$=($T_{BG}$/17.5 K)$^{\beta+4}$. All of these maps have been smoothed to the same angular resolution of 
1$\degr$.

\section{Correlations}

\begin{figure}[!ht]
\resizebox{\hsize}{!}{\includegraphics{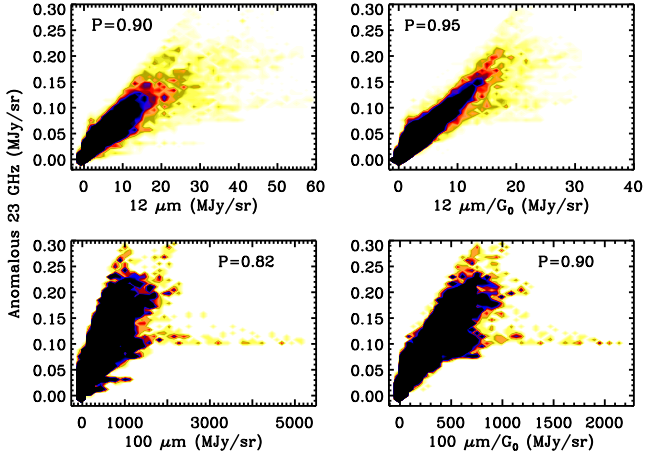}}
\caption{All-sky correlations. Left panels show the anomalous foreground in the 23 GHz WMAP band versus the dust IR emission in the 12 and 100 $\mu$m IRAS bands. In 
the right panels, the dust IR emission has been divided by $G_0$, the intensity of the radiation field. The contrast in point density between yellow and black area is a factor 3. We 
indicate in each panel the value of the Pearson correlation factor $P$.}
\label{correl_ciel} 
\end{figure}

Figure \ref{correl_ciel} shows the all-sky correlation of the 23 GHz anomalous flux with the dust IR emission. The anomalous foreground clearly correlates more strongly with the 12 
$\mu$m band than the 100 $\mu$m (the Pearson correlation factor $P$ is 0.90 and 0.82, respectively). A similar result was obtained by Casassus et al. (2006) towards the LDN
1622 cloud, but here it is the first time that it has been shown to also be true for the entire sky, following the removal of zodiacal light residuals at 12 $\mu$m. The correlation 
is also improved significantly when the dust IR emission is divided by $G_0$ ($P=$ 0.90 to 0.95, in the case of the 12 $\mu$m band\footnote{Using a Monte Carlo method to simulate 
thermal noise in the 12 $\mu$m map, we find that the Pearson coefficent 0.95 differs significantly from 0.90  with a confidence level greater than 99.9\% (using a map containing 
786\,432 pixels).}). This improvement concerns $\sim 60$\% of the sky at 12 $\mu$m. These regions are 1.4 to 1.6 times brighter at 23 GHz than the regions for which the division 
does not improve the correlation. However, in most of the regions where the 
division by $G_0$ does not improve the correlation, it also does not make it poorer. It does only for 5\% of the sky, which could be explained by the uncertainties in the $G_0$ 
values. These correlations show the independence of the anomalous foreground of $G_0$ at 23 GHz and its link with the smallest grains. However, since the all-sky 
correlation is almost as strong with BG emission as with small grains, we are unable to draw firm conclusions at this stage. Across the entire sky, the emissions of PAHs and BGs are known to be 
correlated well. This is no longer true for particular fields, as we now discuss. 

\section{Selected fields}

\begin{figure}[!ht]
\includegraphics[angle=270,width=0.56\textwidth]{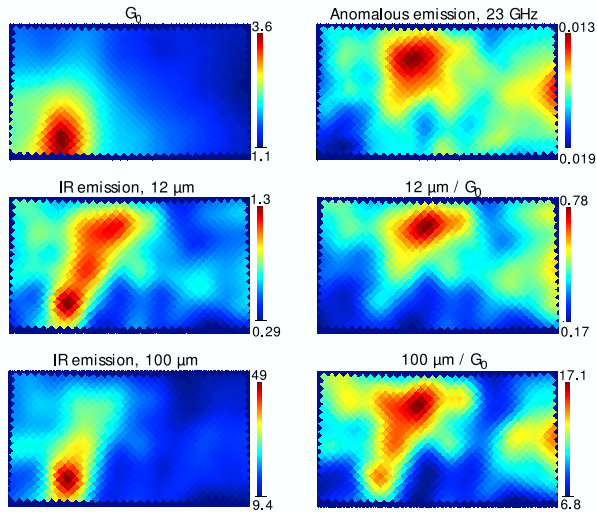}
\caption{Field centred on $l = 343.6 \pm 5.25\degr$ and $b = 22.4 \pm 2.6 \degr$. The 23 GHz, 12 and 100 $\mu$m maps are in MJy/sr.}
\label{region_24_map} 
\end{figure}

\begin{figure}[!ht]
\centerline{
\begin{tabular}{c}
\includegraphics[width=0.4\textwidth]{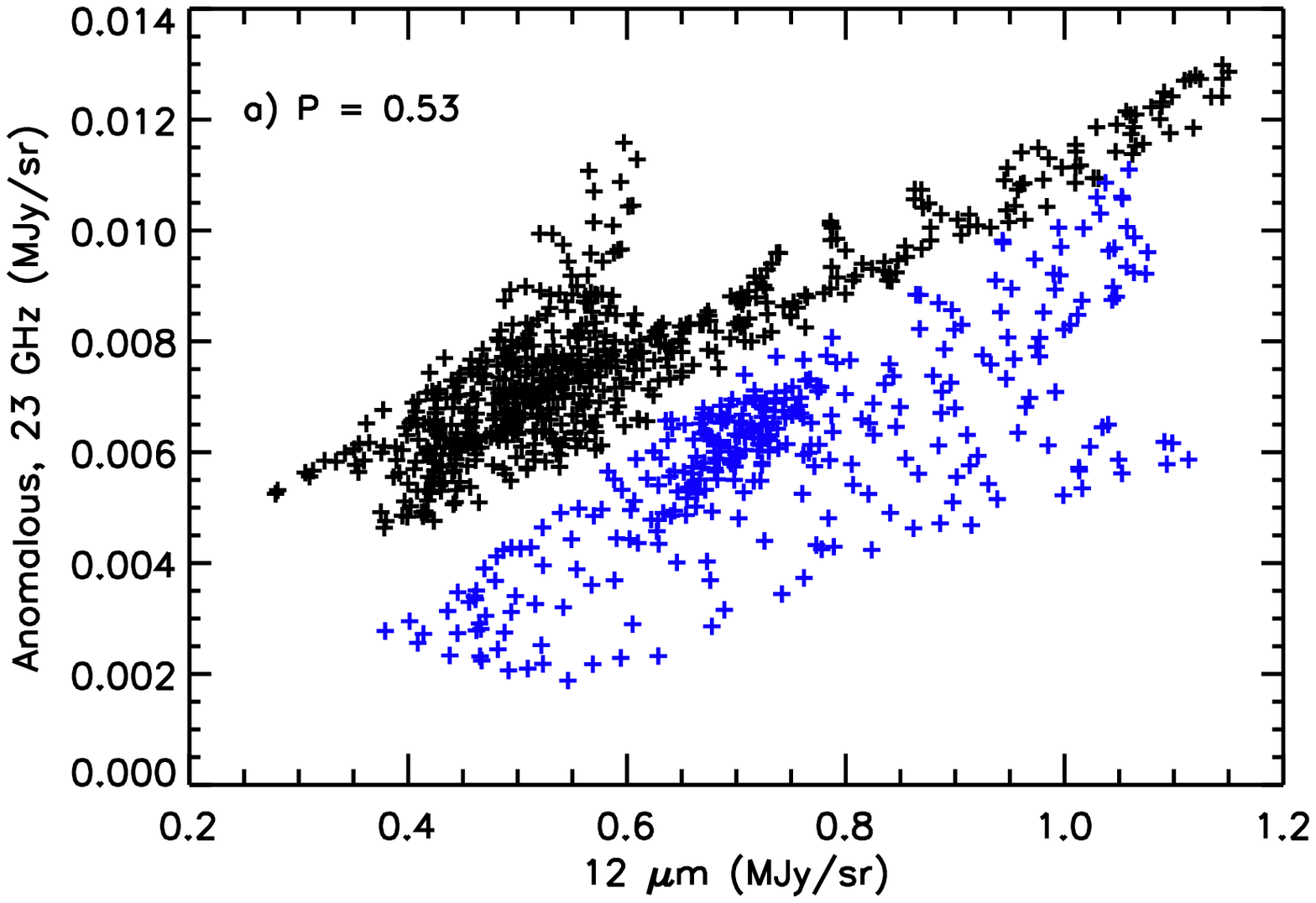} \\
\includegraphics[width=0.4\textwidth]{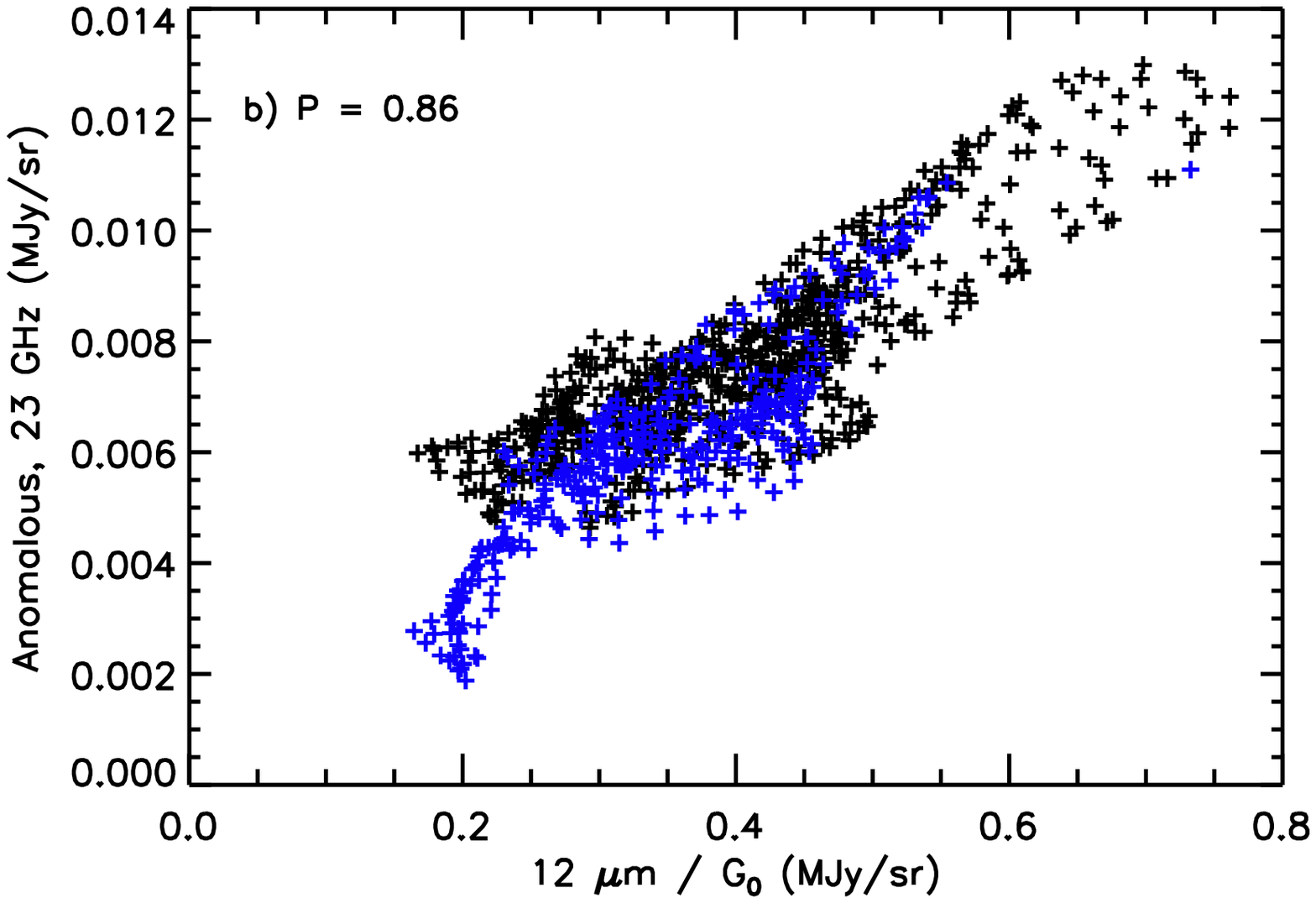}
\end{tabular}}
\caption{Field centred on $l = 343.6 \pm 5.25\degr$ and $b = 22.4 \pm 2.6 \degr$. Correlations between {\it a)} anomalous foreground at 23 GHz and dust 12 $\mu$m emission, 
and {\it b)} anomalous foreground at 23 GHz and dust 12 $\mu$m emission divided by $G_0$. The black crosses show the low-$G_0$ area (median value equal to 1.4) and the blue 
ones the high-$G_0$ area (median equal to 2.3). $P$ is the Pearson's correlation factor.}
\label{region_24} 
\end{figure}

\begin{figure}[!ht]
\resizebox{\hsize}{!}{\includegraphics{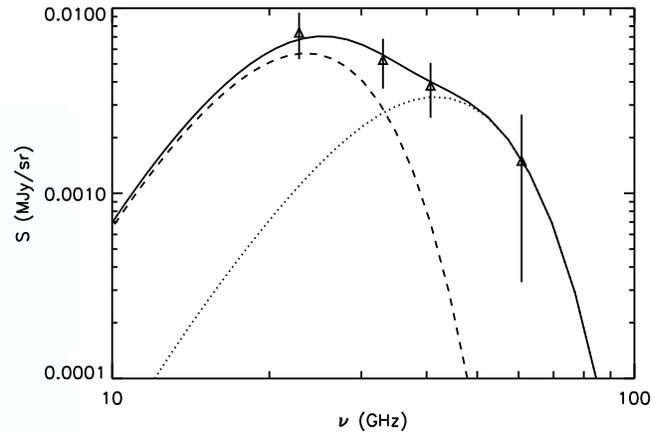}}
\caption{A representative anomalous spectrum of our selected fields (taken from model 4 of MD08). The solid line is the fit with our model (YV09) comprising contributions from the CNM 
(10\%, dashed line) and from the WNM (90\%, dotted line). PAH parameters (see text) are $m=$0.3 D, and $N_{min}=24$ and 48 (CNM and WNM respectively).}
\label{spectre}
\end{figure}

\noindent To test the spinning dust hypothesis further, we searched for fields of a few squared degrees according to the following criteria: location outside of the 
Galactic plane, and bright at both 23 GHz and 12 $\mu$m with $G_0$ variations as large as possible. Searching the sky maps by areas of 5$\degr$ squared, we identified 
27 such fields. Figure \ref{region_24_map} is an example of one of them. The anomalous and dust brightness maps correlate far more tightly when the latter is 
divided by $G_0$\footnote{The improvement in the correlations is significant to a confidence level greater than 99.7\% for the 27 regions.}. The 
correlation plots indeed clearly illustrate two cases (Fig. \ref{region_24}a) corresponding to different values of $G_0$. The difference disappears when the 12 $\mu$m brightness 
is divided by $G_0$ (Fig. \ref{region_24}b), as expected if the anomalous foreground is produced by the emission of spinning PAHs. 

\noindent For 5 of the 27 selected fields, we observe significant spatial variations between the 12 and 100~$\mu$m brightness maps (as shown in Fig. \ref{region_24_map}). In these fields, 
we note that the correlation between the anomalous foreground and the BG 100 $\mu$m/$G_0$ is worse\footnote{The correlation is also poor with the 60 
$\mu$m / $G_0$ and for 3 of them with 25 $\mu$m / $G_0$ ($P = 0.18$ and $P = 0.7$ for the field in Fig. \ref{region_24_map}, respectively).} (Pearson's correlation factor 
$P=0.7$ for the field in Fig. \ref{region_24_map}) than with the smaller grains 12 $\mu$m/$G_0$ ($P=0.86$). This shows that the anomalous foreground is 
correlated well with BG emission, {\it only} if BG emission is well correlated with IR emission characteristics of smaller grains. These results are consistent with spinning dust emission.

\noindent We further test the spinning hypothesis and attempt to constrain the electric dipole moment of PAHs in the selected fields. As discussed by YV09, the brightness 
of spinning PAHs at 23 GHz, $S_{23}$, is given by $N_H\;S_{PAH}\;m^2\epsilon_S$, where $N_H$ is the proton column density, $S_{PAH}$ is the abundance of PAHs, $m$ is a scaling 
factor inferring the electric dipole moment of PAHs, $\mu({\rm D}) = m\times N^{1/2}$ (where $N$ is the number of atoms in the PAH), and $\epsilon_S$ is the rotational luminosity per 
solid angle and per PAH molecule. The PAH IR brightness in the 12 $\mu$m band, $I_{12}$, is proportional to $N_H\,S_{PAH}\,G_0\epsilon_I$, where $\epsilon_I$ is the IR 
luminosity per solid angle and per PAH. The correlation coefficient between anomalous and IR brightness divided by $G_0$ is then $m^2\times\epsilon_S/\epsilon_I$, 
where $\epsilon_S$ depends on the number of carbon atoms in the smallest PAH molecules ($N_{min}$) and the fractions of neutral cold (CNM) and warm (WNM) diffuse gas. In Fig. \ref{spectre}, 
we show a representative fit to the observed anomalous foreground with our model. From the 27 selected fields, we find a mean ratio $S_{23}/(I_{12}/G_0)=1.3\times 
10^{-2}$ with a standard deviation of $4\times10^{-3}$. Our model fits yield $m$ = 0.3 - 0.4 D, $N_{min}=20-60$ and about 10\% of CNM to account for both the 23 GHz and 12 $\mu$m 
emission. These sizes are currently invoked to explain the 3.3 $\mu$m profile in interstellar clouds \citep{Verstraete01, Pech2002} and the $m$-value is in good agreement with laboratory 
measurements for organic molecules (DL98). Thus, the rotational and vibrational emission of PAHs, as in current models, can consistently explain the anomalous and 12 $\mu$m emission 
for plausible properties of PAHs. 

\section{Conclusions}

From an all-sky, degree-scale comparison of the 23 GHz anomalous map with dust IR emission, we have found that the anomalous foreground is well correlated with the 100 $\mu$m IRAS band. 
Using an enhanced set of IRAS maps, we have shown for the first time that the anomalous foreground is correlated with the 12 $\mu$m band across the entire sky and that the correlation is tighter 
than with the 100 $\mu$m flux. This correlation becomes even tighter when the 12 $\mu$m flux is corrected for the intensity of the radiation field $G_0$, indicating that the anomalous 
emission is independent of $G_0$ at 23 GHz on a 1 degree scale. These findings strongly argue in favor of a spinning dust origin to the anomalous foreground. Current 
models predict that the spinning dust emission is dominated by the smallest dust grains (PAHs) carrying the 12 $\mu$m flux and that the corresponding 23 GHz emission is almost 
independent of $G_0$.  From a model fit of both microwave and IR data in selected fields with strong $G_0$ contrast, we deduce the physical properties of PAHs (sizes, electric dipole 
moment) that are in good agreement with results obtained from mid-IR spectroscopy.

\acknowledgements{We thank our referee, Simon Casassus, for his insightful comments that helped in improving the content of this letter. Some of the results in this paper have 
been derived using the HEALPix package \citep{Gorski2005}. This paper used the photoionization code CLOUDY \citep{CLOUDY}.}

\bibliographystyle{aa} 
\bibliography{biblio}

\end{document}